\documentclass[twocolumn,showpacs,floatfix,eqsecnum,prb]{revtex4}

\usepackage[dvips]{epsfig}

\usepackage{float}

\begin{document}

\title{Time Reversal Polarization and a $Z_2$ Adiabatic Spin Pump}

\author{Liang Fu and C.L. Kane}
\affiliation{Dept. of Physics and Astronomy, University of Pennsylvania,
Philadelphia, PA 19104}

\begin{abstract}
We introduce and analyze a class of one dimensional insulating
Hamiltonians which, when adiabatically varied in an appropriate
closed cycle, define a ``$Z_2$ pump".  For an isolated system
a single closed cycle of the pump changes
the expectation value of the spin at each end even when spin orbit
interactions violate the conservation of spin.  A second cycle,
however returns the system to its original state.
When coupled to leads, we show that the $Z_2$ pump functions as a
spin pump in a sense which we define, and transmits a finite, though
non quantized spin in each cycle.
We show that the $Z_2$ pump is characterized by a $Z_2$ topological
invariant that is analogous to the Chern invariant that
characterizes a topological charge pump.
The $Z_2$ pump is closely
related to the quantum spin Hall effect, which is characterized
by a related $Z_2$ invariant.  This work presents an alternative
formulation which clarifies both the physical and mathematical
meaning of that invariant.
A crucial role is played by time reversal symmetry, and we introduce
the concept of the time reversal polarization, which characterizes
time reversal invariant Hamiltonians and signals the presence or
absence of Kramers degenerate end states.  For non interacting
electrons we derive a formula for the time reversal polarization
which is analogous to the Berry's phase formulation of the charge
polarization.  For interacting electrons, we show that abelian
bosonization provides a simple formulation of the time reversal
polarization.   We discuss implications for the quantum spin Hall
effect, and argue in particular that the $Z_2$ classification of the
quantum spin Hall effect is valid in the presence of electron
electron interactions.

\end{abstract}

\maketitle

\section{Introduction}

In recent years, the advent of spintronics has motivated the search
for methods of generating spin currents with little or no
dissipation.  One class of proposals involves designing an adiabatic
pump in which the cyclic variation of some control
parameters results in the transfer of spin across an otherwise
insulating structure\cite{sharma, mucciolo, zhao, shindou}.
Such a spin pump has
been realized in quantum dot structures\cite{watson}.
A second class of proposals involves using the
spin Hall effect to generate a spin current using an electric field
\cite{murakami1,sinova}.
Interest in this approach has been stimulated by the experimental
observation of spin accumulation induced by the spin Hall effect in
doped GaAs structures\cite{kato,wunderlich}.  In these experiments the spin
current is accompanied by a dissipative charge current. This
motivated Murakami, Nagaosa and Zhang\cite{murakami2}
 to propose an interesting class ``spin Hall
insulator" materials which are band insulators that have, according
to a Kubo formula, a large spin Hall conductivity.
However, the spin current which flows in the bulk of these materials is not a
transport current, and can not be simply measured or extracted.  A
crucial ingredient for the generation of transport currents is the
existence of gapless extended edge states.  Such states are generically not
present in the spin Hall insulators\cite{onada}.

Motivated by the spin Hall insulator proposal, we introduced a
 model of graphene in which the symmetry
allowed spin orbit interactions lead to a quantum spin Hall effect\cite{km1,km2}.
A related phase has been proposed for $GaAs$ in the presence of a
uniform strain gradient\cite{bernevig}.
This phase is characterized by a bulk excitation gap and gapless edge
excitations.
In the special case where the spin $S_z$ is conserved, this phase can
be viewed as two copies of the quantum Hall state introduced by
Haldane\cite{haldane}.  The phase persists, however, in the presence
of spin nonconserving interactions as well as disorder\cite{km1,km2,sheng1}.
Time reversal symmetry protects the
gapless edge states when electron interactions are weak, though
strong interactions can open an energy gap at the edge accompanied by
time reversal symmetry breaking\cite{wu,xu}.  We argued that
the quantum spin Hall phase is
 distinguished from a band insulator by a $Z_2$ topological
 index\cite{km2},
 which is a property of the
 bulk system defined on a torus.  We suggested a formula for this index
 in terms of the Bloch wavefunctions.  However, the physical meaning
 of this formula and its relation to the edge states was not explicit.

 When placed on a cylinder (or equivalently a Corbino disk), the quantum
 spin Hall system defines a kind of adiabatic pump as a function of
 the magnetic flux threading the cylinder.  In the case where
 $S_z$ is conserved, advancing the flux by one flux quantum results
 in the transfer of spin $\Delta S_z = \hbar$ from one end of the
 cylinder to the other.  This is a spin pump, whose operation
 is analogous to a charge pump which could be constructed with a quantum
 Hall state on a cylinder.  As envisioned by Thouless and co workers
 in the 1980's\cite{thouless,thoulessniu}, the adiabatic charge pumping process is characterized by a
 topological invariant - the Chern number - which is an integer that
 determines the quantized charge that is pumped in the course of a cycle.
 Equivalently, the Chern number provides a topological
 classification of the two dimensional quantum Hall state\cite{tknn,avron,kohmoto,niu1}.
When $S_z$ is conserved, similar ideas can be used to describe a
quantized adiabatic spin pump\cite{shindou}.

A local conservation law is essential for this type
of topological pumping process.  For a finite
system with closed ends, the eigenstates
before and after a complete cycle must be distinct.  This means that
two energy levels must cross in the course of the cycle.
In the case of the charge pump, that level crossing is protected by
local charge conservation because the two states differ in their
eigenvalue of the charge at each end.
In the absence of a conservation law there will, in
general, be no level crossings, and the system will be in the same
state before and after the cycle.

Unlike charge, spin does not obey a fundamental conservation law, so
unless spin non conserving processes can be made very small it is not
possible to define an adiabatic spin pump that works analogously to
the Thouless charge pump.  Nonetheless, in Ref. \onlinecite{km2}
 we argued that {\it time reversal symmetry}
introduces a conservation law which allows for a topological pumping
process. Specifically, we showed that for a quantum spin Hall state
on a cylinder the eigenstates before and after adiabatic flux
insertion are orthogonal, and can not be connected by any local time
reversal invariant operator.  When a second flux is added, however,
the system returns to its original ground state.  In this sense, the quantum spin
Hall effect defines a ``$Z_2$ pump". In one cycle there is no charge
transferred between the two ends.  Since spin is not conserved, the
two states can not be distinguished by a spin quantum number (though
the expectation value of the spin at the end changes by a non
quantized amount).  The question therefore arises: What is it that is
pumped between the ends of the cylinder?

In this paper we examine this issue carefully and introduce a class
of one dimensional models that exhibit a similar pumping behavior
that is protected by time reversal symmetry.  The ``$Z_2$ spin pump"
is analogous to the quantum spin Hall effect in the same sense that
the charge pump is analogous to the ordinary quantum Hall effect. We
introduce the concept of the {\it time reversal polarization}, a
$Z_2$ quantity which signals whether or not a time reversal invariant
one dimensional system has a Kramers degeneracy associated with its
ends. We show that the change in the time reversal polarization in
the course of an adiabatic cycle is related to a $Z_2$ topological
invariant which distinguishes a $Z_2$ spin pump from a trivial cycle
of an insulator which pumps nothing. This $Z_2$ invariant is
equivalent to the invariant introduced in Ref. \onlinecite{km2} to characterize
the quantum spin Hall effect.  The present work, however, provides an
alternative formulation which clarifies both the physical and
mathematical meaning of the invariant.

We study a family of one dimensional Hamiltonians which have a bulk
energy gap and a length which is much larger than the exponential
attenuation length associated with that gap.  We suppose the Hamiltonian
depends
continuously on a ``pumping parameter" $t$, satisfying the following
properties:
\begin{equation}
H[t+T] = H[t],
\label{cond1}
\end{equation}
\begin{equation}
H[-t] = \Theta H[t] \Theta^{-1},
\label{cond2}
\end{equation}
where $\Theta$ is the time reversal operator.
In the case that the one dimensional system corresponds to a two dimensional
system on a cylinder $t/T$ may be viewed
as the magnetic flux threading the cylinder in units of the flux
quantum.  In the course of the cycle time reversal symmetry is
broken.  However, the second constraint ensures that the system passes
through {\it two} distinct points $t_1^*=0$ and
$t_2^*=T/2$ at which the Hamiltonian is time reversal invariant.
Condition (\ref{cond2}) may be relaxed somewhat, but it is essential that
there exist two distinct time reversal invariant points $t_1^*$ and
$t_2^*$ where (\ref{cond2}) is locally valid.
The existence of two such points plays a crucial role
in the topological classification of the pumping cycle.  In
particular, we will show that pumping cycles in which $H[t_1^*]$ and $H[t_2^*]$
have different time reversal polarization are topologically
distinct from trivial cycles.

We will begin in section II by introducing a simple one dimensional
tight binding model which exemplifies the $Z_2$ spin pump.
This model is closely related to a model of a spin pump
that was recently introduced
by Shindou\cite{shindou} which may be applicable to certain spin 1/2 quantum spin
chains, such as Cu-benzonate and Yb$_4$As$_3$.  This tight binding
model incorporates spin non conserving spin orbit interactions
and provides a concrete illustration of the $Z_2$ pumping effect.

In section III we provide a general formulation of the time reversal
polarization for non interacting electrons.  Our discussion closely
parallels the theory of charge polarization\cite{blount,zak,kingsmith,resta,marzari},
in which the charge
polarization is related to the Berry's phase of Bloch
wavefunctions.  We show how the change in the time reversal
polarization defines a $Z_2$ topological invariant characterizing the
pumping cycle.

In section IV we argue that the notion of time reversal polarization
and the topological classification that follows from it can be
generalized to interacting systems.  We describe an interacting
version of our 1d model using abelian bosonization.  This provides a
different formulation of the time reversal polarization, which is
well defined in the presence of interactions.

In section V we conclude by addressing two issues.  In VA we discuss
the implications of the time reversal polarization for the quantum
spin Hall effect.  We argue that the two dimensional quantum spin Hall phase is a
distinct phase from a band insulator even in the presence of electron
electron interactions.  We then prove that this phase
either has gapless edge excitations or exhibits a
ground state degeneracy associated with time reversal symmetry
breaking at the edge.
We also comment on a proposal by Sheng, Weng, Sheng and Haldane\cite{sheng} to
classify the quantum spin Hall effect in terms of a Chern number
matrix.

In VB we ask whether the $Z_2$ spin pump we have defined can
actually pump spin.  Despite the fact that the isolated pump returns
to its original state after two cycles, we argue that when connected
weakly to leads, the $Z_2$ pump {\it does} pump spin, although the
amount of spin pumped in each cycle is not quantized.

In the Appendix we relate different mathematical formulations of the $Z_2$
topological invariant.  We begin by showing that,
like the Chern invariant, the $Z_2$ invariant can be interpreted as an
obstruction to globally defining wavefunctions, {\it provided} a
constraint relating time reversed wavefunctions is enforced.
We then prove that the invariant derived in this paper is equivalent
to the one introduced in Ref. \onlinecite{km2}.

\section{Tight Binding Model}

In this section we introduce a one dimensional tight binding model of
the $Z_2$ spin pump.  This model is closely related to a model
introduced by Shindou as an adiabatic spin pump\cite{shindou}.  Shindou considered
an antiferromagnetic spin 1/2 Heisenberg chain to which two perturbations which
open a gap in the excitation spectrum are added.  The first term is a
staggered magnetic field $h_{st}$ which locks the spins into a Neel ordered
state.  The second is a staggered component to the
exchange interaction $\Delta J_{st}$, which leads
to a dimerized state.    Interestingly,
Shindou suggested that this model may be relevant to certain $S=1/2$
quantum spin chains, such as Cu-benzonate and Yb$_4$As$_3$, in which
spins reside at two crystallographically inequivalent sites.
He argued that in these systems $h_{st}$ can be
controlled by applying a {\it uniform} magnetic field, and $\Delta J_{st}$
 can be controlled with a uniform electric field.

Shindou showed that a cycle in which $\Delta J_{st}$ and $h_{st}$ are
adiabatically varied defines a topological spin pump, which transfers $S_z=\hbar$
in each cycle.
The topological quantization of
Shindou's pump requires the conservation of $S_z$.  In general,
however, $S_z$ non conserving processes are allowed by symmetry.  In
particular the Dzyaloshinskii-Moriya interaction, ${\bf d} \cdot ( {\bf
S}_1 \times {\bf S}_2$), is allowed, and will inevitably lead to
the violation of $S_z$ conservation.  We will argue, however,
that provided the system retains time reversal invariance when
$h_{st}=0$ this system remains a $Z_2$ spin pump even in the presence
of the Dzyaloshinskii-Moriya interaction.

We study a non interacting electron version of the Shindou model,
where in addition to the spin degree of freedom we allow
charge fluctuations.  Consider a one dimensional
tight binding model with a staggered magnetic field, a
staggered bond modulation as well as a time reversal invariant
spin orbit interaction,
\begin{equation}
H = H_0 +  V_h +  V_t +  V_{so},
\label{ham}
\end{equation}
where
\begin{equation}
H_0 = t_0 \sum_{i,\alpha} \left(c_{i\alpha}^\dagger c_{i+1\alpha} +
c_{i+1\alpha}^\dagger c_{i\alpha}\right),
\end{equation}
\begin{equation}
V_h = h_{st}\sum_{i,\alpha\beta} (-1)^i \sigma^z_{\alpha\beta} c_{i\alpha}^\dagger c_{i\beta},
\end{equation}
\begin{equation}
V_t =  \Delta t_{st}\sum_{i,\alpha} (-1)^i (c_{i\alpha}^\dagger c_{i+1\alpha}
+ c_{i+1\alpha}^\dagger c_{i\alpha})
\end{equation}
and
\begin{equation}
V_{so} = \sum_{i,\alpha,\beta} i \vec e_{so} \cdot \vec \sigma_{\alpha\beta}
(c_{i\alpha}^\dagger c_{i+1 \beta}- c_{i+1\alpha}^\dagger
c_{i\beta}).
\label{hamso}
\end{equation}
Here $\vec e_{so}$ is an arbitrary vector characterizing the spin orbit
interaction. This term
explicitly violates the conservation of $S_z$, playing a role similar
to the Dzyaloshinskii-Moriya interaction in Shindou's model.
We consider an adiabatic cycle in which
\begin{equation}
(\Delta t_{st},h_{st}) = (\Delta t^0_{st} \cos ( 2\pi t/T), h^0_{st} \sin (2\pi
t/T)).
\end{equation}
Since $V_h$ is odd under time reversal, while $V_t$ is even, condition
(\ref{cond2}) is clearly satisfied.  At $t=0$ and $T/2$ the Hamiltonian is
time reversal invariant.

\begin{figure}
 \centerline{ \epsfig{figure=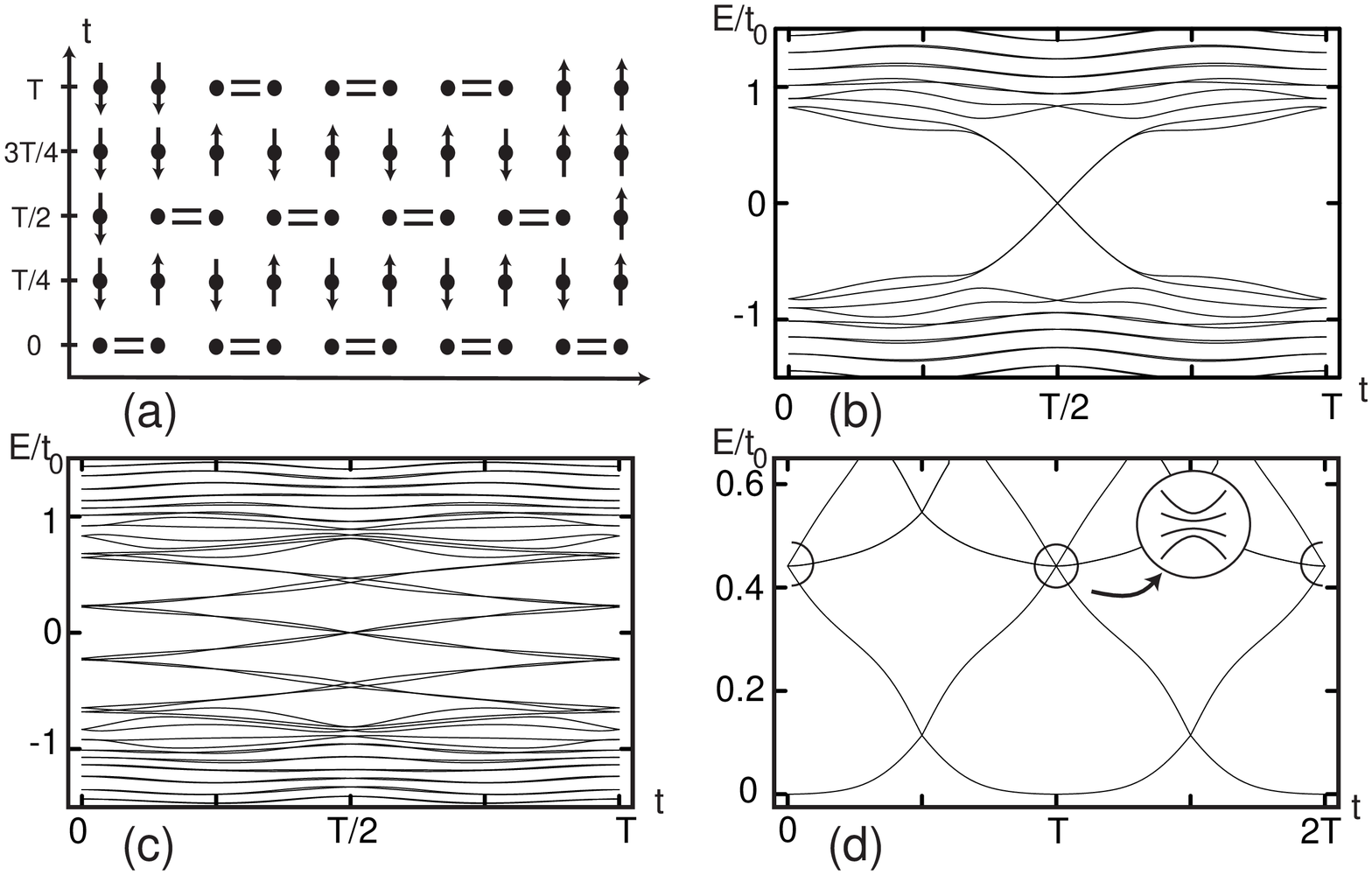,width=3.5in} }

 \caption{
 (a) Schematic representation of the groundstate of Eqs.
 (\ref{ham}-\ref{hamso})
 for various $t$.  The groundstates at the time reversal invariant points
 $t=0$ and $t=T/2$ are distinguished by the presence of Kramers degenerate
 end states.  (b) Single particle energy levels $E_n(t)$ for a 24 site chain with
 $\Delta t^0_{st}/t_0 = .4$, $h^0_{st}/t_0 = .8$ and $\vec e_{so}/t_0 = .1 \hat y$.
 (c) Single particle energy levels $E_n(t)$ for a 24 site chain with
 12 extra sites added at each end.  (d) Low energy many body energy levels
 associated with one end of the chain.  The degeneracy at $t=T/2$ and
 $t=3T/2$ is protected by time reversal symmetry.  The inset shows
 how electron electron interactions lifts the degeneracy at $t=0$,
 $t=T$ and $t=2T$.
  }
 \end{figure}

In Fig. 1(a) we depict groundstates in the
strong coupling limit at representative points along the cycle. At $t=T/4$ and
$t=3T/4$ $V_h$ dominates and locks the spins into a Neel ordered
state.  At $t=0$ and $t=T/2$, $V_t$ dominates, and the system is
dimerized with singlet pairs of electrons occupying alternate bonds.
Importantly, the groundstate
at $t=T/2$ is distinguished from the groundstate at $t=0$ by the
presence of {\it unpaired} spins at each end.

When $V_{so}=0$ $S_z$ is
conserved, and this model describes a spin pump.  In this case
$V_h+V_t$ can be decomposed into
two independent periodic potentials which lock the densities of
the up and down spin particles and slide in opposite directions
as a function of $t$.  As $t$ evolves from $0$ to $T$, the
periodic potentials slide by one lattice
constant.  Provided there is ``space" for the added spin at the ends,
spin $\hbar$ will accumulate at the end following each cycle.

We wish to understand how this spin pump is modified when $V_{so}\ne
0$, so that $S_z$ is not conserved.  In Fig. 1(b) we plot the single
particle energy levels for a $24$ site chain as a function of $t$ for
nonzero $V_{so}$.
The bulk energy gap can be clearly seen with continuum states above
and below.  The energy levels that cross the gap are end states.
Each line consists of two states which are localized at opposite
ends.   The crossing of the end states at $t=T/2$ will play a
critical role in what follows.   When $V_{so}=0$ the degeneracy at $T/2$ is
protected by spin conservation because the two states at each edge
have $S_z=\pm \hbar/2$.  Nonzero $V_{so}$ does {\it not} lift the
degeneracy provided the Hamiltonian remains time reversal
invariant at $T/2$.  The two end states form a Kramers doublet whose
degeneracy can not be broken by any time reversal invariant
perturbation.

Because of the level crossing at $t=T/2$ it is clear that a system
which starts in the ground state at $t=0$ will be in an excited state
at $t=T$.   However, since the end states merge with the continuum
the excitation will not be localized near the edge, and bulk
particle-hole pairs will be excited.  This is because there is no
``space" to put the excitations at the ends.  In section VB we will
discuss the effect of connecting this pump to reservoirs which allow
the end states to be ``emptied" without exciting bulk particle-hole
pairs. For the purpose of this section, however, we will study the
operation of an isolated pump by adding several sites at the ends of
the chain for which $V_h$ and $V_t$ vanish.  This introduces
additional
midgap states localized at each end, allowing the cycle to proceed
without generating bulk excitations.

Fig. 1(c) shows the energy levels as a function of $t$ with the extra
sites added at each end.  There are now several midgap states at
each end.  Since all of the midgap states are localized at one end or
the other, the low energy excitations of the system can be
factorized as a product of excitations at each of the two ends.

In Fig. 1(d)
we plot for $0<t<2T$ the energies of the lowest few many body
eigenstates associated with a single end,
obtained by considering particle hole excitations built from the
single particle states localized at that end.  Though this picture was
computed for non interacting electrons, it is clear that the Kramers
degeneracy of the groundstate at  $t=T/2$ and $3T/2$ will be robust to
the addition of
electron electron interactions.  The first excited state at $T=0$,
$T$ and $2T$ in Fig. 1(d) is four fold degenerate (the middle
 level coming into that point is doubly degenerate).
This degeneracy, however, is an
artifact of non interacting electrons.  The degeneracy is present
because there are four ways of making particle hole excitations with
two pairs of Kramers degenerate states.  Electron electron interactions,
however, will in general split this degeneracy, as shown in the
inset, so there will be no level crossing at $t=T$.

We thus conclude that when the isolated pump starts in its ground
state at $t=0$, it arrives in an excited state after one complete
cycle at $t=T$.  After a second cycle, however, at $t=2T$ the system
returns to its original state.  For this reason, we call it a ``$Z_2$
pump".  It is possible that by coupling to other degrees of freedom
an {\it inelastic} process (such as emitting a phonon)
could cause the excited state to relax back to the ground state.
Nonetheless, there is an important distinction between this adiabatic
process which generates an excited state and one that does not.
In section VB we will return to this issue when we discuss connecting
the pump to leads.
The nontrivial operation of a single cycle depends critically on the time
reversal symmetry at $t=T/2$.  Breaking time reversal symmetry at
that point leads to an avoided crossing of the energy levels, so
that the system returns adiabatically to its original state at $t=T$.

From the point of view of the end states, the non trivial pumping
effect arises because there exist Kramers degenerate end states at
$T=T/2$, but not at $T=0$.  In the next section we show that this
property is determined by the topological structure of the {\it bulk}
Hamiltonian, $H(t)$.

\section{Time Reversal Polarization and $Z_2$ Invariant}

In this section we introduce the time reversal polarization
for non interacting electrons and show that changes in it define
a topological invariant.  Our discussion will parallel the
theory of charge polarization in insulators\cite{blount,zak,kingsmith,resta,marzari}.
In order to establish
this connection and to define our notation we will therefore begin by
reviewing that theory, which relates the charge polarization to the average
center of Wannier orbitals, which in turn are related to the Berry's phase
of the Bloch wavefunctions.  We next
consider the role of Kramers' degeneracy in time reversal invariant
systems and define a corresponding time reversal polarization in terms
of the difference between the Wannier centers of Kramers degenerate bands.
Finally, we show that the {\it change} in the time reversal polarization
between $t=0$ and $t=T/2$ of the pumping cycle defines a $Z_2$
topological invariant which distinguishes a nontrivial $Z_2$ pump
from a trivial cycle.

\subsection{Review of theory of charge polarization}

Consider a one dimensional system with lattice constant $a=1$,
length $L = N_c$ with periodic boundary conditions and $2N$ occupied bands.
The normalized eigenstates for the $n$'th band
can then be written in terms of cell
periodic Bloch functions as
\begin{equation}
|\psi_{n,k}\rangle = {1\over\sqrt{N_c}}e^{ikx}|u_{n,k}\rangle.
\end{equation}
We may define Wannier functions associated with each unit cell
associated with lattice vector $R$ as
\begin{equation}
|R,n\rangle = {1\over {2\pi}} \int dk e^{-i k(R-r)} |u_{k,n}\rangle.
\end{equation}
The Wannier functions are not unique because they depend on a gauge
choice for $|u_{k,n}\rangle$.  In addition to changing the phases of
the individual wavefunctions, the wavefunctions can be mixed by a
general $U(2N)$ transformation of the form
\begin{equation}
|u_{k,n}\rangle \rightarrow \sum_m U_{nm}(k) |u_{k,m}\rangle.
\label{u2n}
\end{equation}
After this transformation, $|u_{k,n}\rangle$ need no longer
be the individual eigenstates of the Hamiltonian,
but rather should be interpreted as
basis vectors spanning the space spanned by the $2N$ occupied eigenstates.
The Slater determinant of the $2N$ wavefunctions is unchanged up to a phase.

Marzari and Vanderbilt\cite{marzari} have provided a prescription
for choosing $U_{nm}(k)$ to optimally localize the Wannier wavefunctions.
Here, however, we are
concerned with the total charge polarization, which is insensitive to
the details of $U_{nm}(k)$.  The polarization is given by the sum
over all of the bands of the center of charge of the Wannier states
associated with $R=0$, and may be written\cite{blount,zak}
\begin{equation}
P_\rho = \sum_n \langle 0,n|r|0,n\rangle = {1\over {2\pi}}
\oint dk \ {\cal A}(k).
\label{prho}
\end{equation}
where the $U(1)$ Berry's connection is given by
\begin{equation}
{\cal A}(k) = i\sum_n\langle u_{k,n}|\nabla_k|u_{k,n}\rangle.
\end{equation}
The integral is over the Brillouin zone from $k=-\pi$ to $\pi$.
If we require that the wavefunction $|\psi_{n,k}\rangle$ be defined
continuously in the reduced zone scheme, so that
$|\psi_{n,-\pi}\rangle = |\psi_{n,\pi}\rangle$ then ${\cal A}(-\pi) =
{\cal A}(\pi)$, and the integral may be considered to be on a {\it
closed} loop, despite the fact that $|u_{n,k}\rangle$ is discontinuous
from $-\pi$ to $\pi$\cite{zak}.  Under $U(2N)$ a transformation which preserves this
continuity $P_\rho$ is invariant up to a lattice constant.
For a transformation  in which the $U(1)$ phase of $U_{mn}(k)$
advances by $2\pi m$ when $k$ advances around the Brillouin zone
$P_\rho \rightarrow P_\rho + m$.  This reflects the fact that the
polarization can only be defined up to a lattice vector.

{\it Changes} in the polarization induced by a continuous change in the
Hamiltonian $H[t]$ are, however, well defined.  Thus, if the wave
functions $|u_{k,n}(t)\rangle$ are defined {\it continuously} between $t_1$
and $t_2$ for all $k$ in the Brillouin zone, then we may write
\begin{equation}
P_\rho[t_2]-P_\rho[t_1] = {1\over {2\pi}}\left[
\oint_{c_2}dk\  {\cal A}(t,k) -\oint_{c_1}dk \ {\cal A}(t,k)
\right],
\end{equation}
where $c_{1(2)}$ is the loop $k = -\pi$ to $\pi$ for fixed
$t=t_{1(2)}$.
Using Stokes theorem, this can be written as an integral of the Berry
curvature
\begin{equation}
 {\cal F}(t,k)
=i \sum_n \left(\langle \nabla_t u_{k,n}(t) |\nabla_k u_{k,n}(t)\rangle -
c.c\right)
\end{equation}
over the surface $\tau_{12}$ of the cylinder spanned by $k$ and $t$
bounded by $c_1$ and $c_2$:
\begin{equation}
P_\rho[t_2]-P_\rho[t_1] = {1\over {2\pi}} \int_{\tau_{12}} dt dk\  {\cal
F}(t,k).
\label{prho2}
\end{equation}

For a periodic cycle $H[t+T]=H[t]$, the change in the polarization
over one cycle, $P_\rho(T)-P_\rho(0)$ is given by the integral in (\ref{prho2})
over the entire torus defined by $t$ and $k$.  This quantity is
an integer and defines the first
Chern number associated with the wavefunction
$|u_{k,n}(t)\rangle$ on the torus.  The Chern number characterizes
the charge pumped in each cycle.   For a cycle which satisfies the
time reversal constraint in Eq. (\ref{cond2}), ${\cal F}(-t,-k) = - {\cal
F}(t,k)$, so the Chern number is equal to zero.

\subsection{Time reversal Polarization for Kramers degenerate bands}

Consider now a time reversal invariant system.  The time reversal
operator has the form
\begin{equation}
\Theta = e^{i \pi  S_y/\hbar} K,
\end{equation}
where $S_y$ is spin operator and $K$ is complex conjugation.
Since $\Theta^2=-1$ for spin 1/2 electrons, it follows from Kramers' theorem
that every Bloch state at wavevector $k$ is degenerate with a
time reversed Bloch state.  Therefore, the energy bands come in
pairs, which are degenerate at the two time
reversal invariant points $k^* = 0$ and $\pi$, as shown in Fig. 2.
Note that in the presence of spin orbit interactions these bands can
{\it not} be labeled with spin quantum numbers.

\begin{figure}
 \centerline{ \epsfig{figure=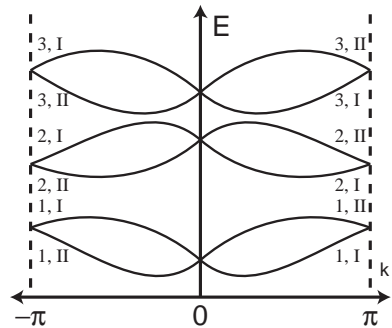,width=2in} }
 \caption{Schematic one dimensional band structure with spin orbit interactions.
 The energy bands come in time reversed pairs which are
 degenerate at $k=0$ and $k=\pi$. }
 \end{figure}

In section IIIA we related the charge polarization as the sum of the
Wannier centers of all of the bands.  Kramer's theorem guarantees,
however that the Wannier states come in Kramer's degenerate pairs, in
which each pair has the same center.  The idea is therefore to keep
track of the center of {\it one} of the degenerate Wannier states per
pair by defining a ``partial polarization".
This will contain more information than Eq. (\ref{prho}), which is the
sum over both states.

For simplicity we assume that there are no degeneracies other than
those required by time reversal symmetry.  Therefore, the $2N$ eigenstates
may be divided into $N$ pairs which satisfy
\begin{eqnarray}
|u_{-k,\alpha}^I\rangle  =& e^{i\chi_{k,\alpha}} \Theta |u_{k,\alpha}^{II}\rangle
\nonumber \\
|u_{-k,\alpha}^{II}\rangle =& - e^{i\chi_{-k,\alpha}} \Theta |u_{k,\alpha}^I \rangle,
\label{timerev}
\end{eqnarray}
where $\alpha=1, ..., N$.  The second equation follows from the first, along with the
property $\Theta^2=-1$.  As shown in Fig. 2 these bands are defined
continuously at the degeneracy points $k^*=0, \pi$.  This
representation is not invariant under the general $U(2N)$
transformation (\ref{u2n}).  However, that invariance will be
restored below.

We define Wannier states associated with these two sets of bands along
with the corresponding Wannier centers.  By analogy with (\ref{prho})
the partial polarization associated
with one of the categories $s = I$ or $II$ may then be written
\begin{equation}
P^s = {1\over {2\pi}} \int_{-\pi}^{\pi} dk {\cal A}^s(k),
\label{ps}
\end{equation}
where
\begin{equation}
{\cal A}^s(k) = i \sum_\alpha\langle
u_{k,\alpha}^s|\nabla_k|u_{k,\alpha}^s\rangle.
\end{equation}
The partial polarizations are clearly invariant (up to a lattice
translation) under changes in the phases of $|u_{k,\alpha}^I\rangle$ and
$|u_{k,\alpha}^{II}\rangle$.  However, they appear to depend on the arbitrary
choice of the labels $I$ and $II$ assigned to each band.
We now show that the partial polarizations (\ref{ps}) can be written in a form that
is invariant under
a general $U(2N)$ transformation of the form (\ref{u2n}).
To make this invariance
explicit for $P_I$ we treat the portions of the integral for positive
and negative $k$ separately,
\begin{equation}
P^I = {1\over {2\pi}} \int_0^{\pi} dk \left[
{\cal A}^I(k) + {\cal A}^I(-k) \right].
\end{equation}
For the second term we use the time reversal constraint (\ref{timerev}) along with
the fact that $\langle \Theta u^{II}_{k,\alpha}|\nabla_k|\Theta u^{II}_{k,\alpha}\rangle = -
\langle u^{II}_{k,\alpha}|\nabla_k|u^{II}_{k,\alpha}\rangle$ to write
\begin{equation}
{\cal
A}^I(-k) = {\cal A}^{II}(k) - \sum_\alpha \nabla_k \chi_{k,\alpha}.
\end{equation}
It then follows that
\begin{equation}
P^I = {1\over{2\pi}} \left[ \int_0^{\pi} dk  {\cal A}(k)
 - \sum_\alpha \left(\chi_{\pi,\alpha}-\chi_{0,\alpha}\right)\right].
 \label{pi1}
\end{equation}
The first term is expressed in terms of the Berry's connection
${\cal A} = {\cal A}^I + {\cal A}^{II}$.  However, since the path of
integration is not closed, the second term
is necessary to preserve gauge invariance.  The second term,
can be rewritten in a suggestive manner by introducing the $U(2N)$ matrix
which relates the time reversed wavefunctions,
\begin{equation}
w_{mn}(k) = \langle u_{-k,m}|\Theta|u_{k,n}\rangle.
\end{equation}
In the representation (\ref{timerev}) $w_{mn}$ is a
direct product of two by two matrices with $e^{i\chi_{k,\alpha}}$ and
$-e^{i\chi_{-k,\alpha}}$ on the off diagonal.  At $k=0$ and $k=\pi$
$w_{mn}$ is {\it antisymmetric}.  An antisymmetric
matrix may be characterized by its Pfaffian, whose square is equal to
the determinant.  We then find that
\begin{equation}
{{\rm Pf}[w(\pi)] \over{ {\rm
Pf}[w(0)]}} = \exp[i \sum_\alpha (\chi_{\pi,\alpha}-\chi_{0,\alpha})].
\end{equation}
Thus, the second term in (\ref{pi1}) can be expressed in terms of ${\rm
Pf}[w]$.  This leads to
\begin{equation}
P^I = {1\over{2\pi}} \left[\int_0^{\pi} dk {\cal A}(k)
+i {\rm log}\left({{\rm Pf}[w(\pi)] \over {{\rm
Pf}[w(0)]}}\right)\right].
\label{pi2}
\end{equation}
Using the identity ${\rm Pf}[XAX^T]={\rm Det}[X]{\rm Pf}[A]$ it
can be shown that under the $U(2N)$ transformation (\ref{u2n})
${\rm Pf}[w]  \rightarrow {\rm Pf}[w]
{\rm Det}[U]$.  Both terms in (\ref{pi2}) are thus clearly
$SU(2N)$ invariant.  Moreover, under a $U(1)$
transformation the two terms compensate one another, so $P^I$ is
$U(2N)$ invariant.  Like the charge polarization (\ref{prho}), $P^I$ is only
defined modulo a lattice vector.   This is reflected in the ambiguity
of the imaginary part of the log in (\ref{pi2}) as well as the dependence of
gauge transformations where the phase of $|u_{k,n}\rangle$
advances by $2\pi$ for $0<k<\pi$.

A similar calculation can be performed for $P^{II}$, and it is clear
from time reversal symmetry that $P^{II}=P^I$ modulo an integer,
reflecting the Kramer's pairing of the Wannier states.
From (\ref{prho}) and (\ref{ps}) the charge polarization is given by the sum of
the two partial polarizations:
\begin{equation}
P_\rho = P^I + P^{II}.
\end{equation}
We now define the {\it time reversal polarization} as the
difference:
\begin{eqnarray}
P_\theta &= P^I - P^{II}  \nonumber \\
&= 2 P^I - P_\rho.
\end{eqnarray}
This then has the form,
\begin{equation}
P_\theta = {1\over{2\pi}}\left[\int_0^\pi dk {\cal A} -\int_{-\pi}^0 dk
 {\cal A}
+  2i {\rm
log}\left({{\rm Pf}[w(\pi)] \over {{\rm
Pf}[w(0)]}}\right)\right].
\label{pth1}
\end{equation}
This may be written more compactly in terms of $w_{mn}$ as
\begin{equation}
P_\theta = {1\over{2\pi i}} \left[\int_0^\pi dk {\rm Tr}[w^\dagger\nabla_k w] -
2 {\rm log} \left({{\rm Pf}[w(\pi)]\over{{\rm Pf}[w(0)]}}\right)\right].
\label{pth2}
\end{equation}
This can further be simplified by noting that the first term gives
the winding of the $U(1)$ phase of $w_{mn}$ between $0$ and
$\pi$.  Thus,
\begin{equation}
P_\theta = {1\over{2\pi i}}\left[
\int_0^\pi dk \nabla_k {\rm log} {\rm Det}[w(k)]
 - 2 {\rm log}\left({{\rm Pf}[w(\pi)] \over
{{\rm Pf}[w(0)]}}\right)\right].
\label{pth3}
\end{equation}
Since ${\rm Det}[w] = {\rm Pf}[w]^2$ this
quantity is an integer, and the due to the ambiguity of the log, this
integer is only defined modulo 2.  Even and odd integers are distinct,
however, and determine whether ${\rm Pf}[w(k)]$ is on the same
branch or opposite branch of
$\sqrt{{\rm Det}[w(k)]}$ at $k=0$ and $\pi$.  An alternative way of
writing it is thus,
\begin{equation}
(-1)^{P_\theta} =
{\sqrt{{\rm Det}[w(0)]}\over{{\rm Pf}[w(0)]}}
 {\sqrt{{\rm Det}[w(\pi)]}\over{{\rm Pf}[w(\pi)]}},
 \label{pth4}
\end{equation}
where the branches of $\pm\sqrt{{\rm Det}[w]}$ are chosen such that the
branch chosen at $k=0$ evolves continuously along the path of integration
in (\ref{pth3}) into the branch chosen at
$k=\pi$ - eliminating the ambiguity of the square root.

Eqs. (\ref{pth1}-\ref{pth4}) are among the
 principle results of this paper, and can
be regarded as a generalization accounting for time reversal symmetry
of the Berry's phase formulation of the charge polarization\cite{zak}.
The $Z_2$ time reversal polarization $P_\theta$ defines two
distinct polarization states.  In the next section we will argue that
the value of $P_\theta$ is related to the presence or the absence of
a Kramers degenerate state at the end of a finite system.
 As is the case for $P_\rho$, the value of $P_\theta$ is not meaningful
by itself, because a gauge transformation
$|u_k^I\rangle \rightarrow e^{ik}|u_k^I\rangle $
changes its value.  Equivalently, the presence or absence of a
Kramers degeneracy at the end can not be determined from the state in
the bulk, since it will depend on how the crystal is terminated.
Nonetheless, the two values of $P_\theta$ are topologically
distinct in the sense that the value of $P_\theta$ can not be altered by a
continuous change in the Hamiltonian which preserves time reversal
symmetry.  However, in the next section we will argue
that an adiabatic change in the
Hamiltonian which preserves time reversal symmetry at the end
points -
but not in between - leads to a well defined change in $P_\theta$.
This change defines a topological classification of distinct pumping
procedures.

\subsection{$Z_2$ Invariant}

In the previous subsection we focused on a time reversal invariant
Hamiltonian, which occurs at $t=0$ and $T/2$ in our pumping cycle.
We now consider the continuous evolution of the Hamiltonian through
the cycle and show that the change in the time reversal polarization
which occurs in {\it half} the cycle defines a $Z_2$ topological
invariant which distinguishes a $Z_2$ spin pump from a trivial cycle.

This physical meaning of this invariant
is easiest to see pictorially by considering the
shift in the Wannier centers in the course of one cycle.  Fig. 3(a)
depicts the centers of the occupied Wannier orbitals
as a function of $t$.  At $t=0, T/2$ and
$T$, time reversal symmetry requires that the Wannier states come in
time reversed pairs.  However, in going from $t=0$ to $t=T/2$ the
Wannier states ``switch partners".  In this process, the time
reversal polarization, which tracks the difference between the
positions of the time reversed Wannier states changes by one.  In
addition, this switching results in the appearance of an unpaired
occupied Wannier state at each end.  Since the Wannier states come in
pairs, there must be twofold Kramers degeneracy
associated with each end - resulting in a total degeneracy of four.

When the system evolves from $t=T/2$ to $t=T$, there is another
switch, and the time reversal polarization returns to its original
value.  However, since $H[t] = \Theta H[T-t]\Theta^{-1}$, the system
with open ends does not return to its original state at $t=0$ but its
ends are in an excited state because of the level crossing at $t=T/2$.

\begin{figure}
 \centerline{ \epsfig{figure=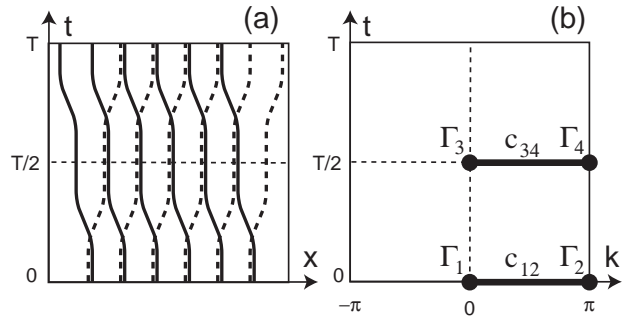,width=3.2in} }
 \caption{(a) Schematic diagram showing the evolution of the centers of the
 time reversed pairs of Wannier states as a function of $t$.  Between $t=0$ and
 $t=T/2$ the Wannier states ``switch partners", resulting in the appearance of
 unpaired Wannier states at the end.  (b) The torus defined by $k$ and $t$,
 with the four time reversal invariant points $\Gamma_i$ connected by paths
 $c_{12}$ and $c_{34}$. }
 \end{figure}

We now relate the occurrence of this non trivial pumping cycle to a
topological property of the {\it bulk ground state} as a function of
$t$.  We thus consider the change in the time reversal polarization,
$P_\theta(t)$
between $t=0$ and $t=T/2$.  Note that though $P_\theta$ is not gauge
invariant, the {\it change} in $P_\theta$ is gauge invariant.  This
difference
\begin{equation}
\Delta = P_\theta(T/2) - P_\theta(0) \ {\rm mod} \ 2
\label{delta1}
\end{equation}
defines a $Z_2$ topological invariant which characterizes the mapping from the
torus defined by $k$ and $t$ to the wavefunctions:
$|u_{k,n}(t)\rangle$.  From (\ref{pth4}) we may write this invariant as
\begin{equation}
(-1)^\Delta = \prod_{i=1}^4 {\sqrt{{\rm Det}[w(\Gamma_i)]}\over
{{\rm Pf}[w(\Gamma_i)]}}.
\label{delta2}
\end{equation}
Here $\Gamma_i$ are the four ``time reversal invariant points" on the
torus shown in Fig. 3(b).  The branches of the square root are chosen
as in (\ref{pth4}) by continuously evolving $\sqrt{{\rm
Det}[w(k,t)]}$ along the paths $c_{12}$ and $c_{34}$.
In order to apply this formula,
it is crucial for the wavefunctions to be defined
{\it continuously} on the torus.
It is always possible to find such smoothly defined
wavefunctions via a transformation of the form (\ref{u2n})
because the Chern number - which is the obstruction to
doing so - is equal to zero.

In the Appendix we will relate different mathematical formulations
of this invariant.  We will first
show that it can be interpreted as an obstruction to
defining continuous wavefunctions provided an additional constraint
relating the wavefunctions at time reversed points is enforced.  This
leads to a different formula for the invariant, which can be
expressed in terms of the Berry's curvature ${\cal F}$ and the
Berry's connection ${\cal A}$.  We will then prove that
(\ref{delta2}) is equivalent to the formula for the invariant
introduced in Ref. \onlinecite{km2}.

\section{Electron Interactions and Bosonization}

The preceding discussion has focused on non interacting electrons.
An important question is therefore whether or not these ideas apply
to interacting systems.   The presence or absence
of a groundstate Kramers degeneracy associated with the ends of a finite
interacting time reversal invariant system is
clearly a well posed yes or no question.  This suggests that the time
reversal polarization is a well defined quantity - at least for non
fractionalized phases for which the groundstate with
periodic boundary conditions is non degenerate.  Therefore, we
believe the topological distinction of the $Z_2$ pump is
still present with interactions.

Calculating the time reversal polarization for interacting electrons is more
subtle than for non interacting electrons.  One possible approach
would involve characterizing the {\it entanglement entropy}, as in
Ref. \onlinecite{ryu}, which is sensitive to the presence or absence of end states.
In this section we adopt a simpler approach by
studying an interacting version of the model
introduced in section II using abelian bosonization.  We find
that bosonization provides a natural description of the time reversal
polarization.

We begin with a continuum version of (\ref{ham}) described by the Hamiltonian density,
\begin{equation}
H = \psi^\dagger\left( i v_F\tau^z\partial_x + h_{st} \tau^x \sigma^z + \Delta
t_{st} \tau^y + i \vec e_{so}\cdot \vec \sigma \tau^z \right) \psi.
\end{equation}
Here $\psi_{a\alpha}$ is a four component field, where
the left and right moving fields $a=L,R$ are specified by
the eigenvalues of $\tau^z_{ab}$
and the spin $\alpha = \uparrow\downarrow$ by $\sigma^z_{\alpha\beta}$.
We now bosonize according to
\begin{equation}
\psi_{a\alpha} = {1 \over \sqrt{2\pi x_c}} e^{i \phi_{a\alpha}}.
\end{equation}
where $x_c$ is a short distance cutoff.
Define charge/spin variables so that $\phi_{\uparrow/\downarrow\alpha} =
\phi_{\rho\alpha} \pm \phi_{\sigma\alpha}$, and charge/current
variables (with $\mu = \rho,\sigma$) as $\phi_{\mu L/R} = \varphi_\mu
\pm\theta_\mu$.  These obey $[\partial_x \theta_\mu(x),\varphi_{\mu'}(x')] = i
(\pi/2)
\delta_{\mu\mu'}\delta(x-x')$.

The bosonized Hamiltonian then has the form (\ref{ham}) with
\begin{equation}
H_0 = {v_F\over {4\pi}}\left[(\partial_x\varphi_\rho)^2+(\partial_x\theta_\rho)^2+
(\partial_x\varphi_\sigma)^2+
(\partial_x\theta_\sigma)^2\right],
\end{equation}
\begin{equation}
V_h = {h_{st}\over{2\pi x_c}} \sin 2\theta_\rho \sin 2\theta_\sigma,
\end{equation}
\begin{equation}
V_t = {\Delta t_{st}\over{2\pi x_c}} \sin 2\theta_\rho
\cos 2\theta_\sigma,
\end{equation}
and
\begin{equation}
V_{so} = {e_{so}^z\over\pi} \partial_x \varphi_\sigma +
{e_{so}^x\over{2\pi x_c}}
\sin 2\theta_\sigma \cos 2\varphi_\sigma + {e_{so}^y\over{2\pi x_c}}
 \sin 2\theta_\sigma \sin
2\varphi_\sigma.
\end{equation}
In the absence of the spin orbit term, the spin sector of this
Hamiltonian (when $\theta_\rho$ is pinned at $\pi/4$)
is equivalent Shindou's model\cite{shindou}.  This Hamiltonian describes
an insulating phase in which both $\theta_\rho$ and
$\theta_\sigma$ are pinned.

First focus on
the case $h_{st}=0$ where the Hamiltonian is time reversal invariant.
If we choose a gauge such that
$\Theta \psi \Theta^{-1} = \tau^x \sigma^y \psi^*$, the behavior of these
operators under time reversal can be deduced:
 \begin{eqnarray}
 &\Theta \theta_\rho \Theta^{-1} = \theta_\rho, \ \ \
 &\Theta \varphi_\rho \Theta^{-1} = -\varphi_\rho,  \nonumber\\
 &\Theta \theta_\sigma \Theta^{-1} = -\theta_\sigma, \ \ \
 &\Theta \varphi_\sigma \Theta^{-1} = \varphi_\sigma + \pi/2.
 \end{eqnarray}
The time reversal invariance of the Hamiltonian when $h_{st}=0$ can
easily be verified.  It is now straightforward to consider time
reversal invariant interaction terms, such as
$(\partial_x\theta_\rho)^2$, $(\partial_x\theta_\sigma)^2$,
$\cos 4\theta_\rho$, $\cos 4\theta_\sigma$, $\cos 4\varphi_\sigma$, etc.
Provided these interaction terms (as well $V_{so}$ defined above) are
not too large, the system will retain its bulk gap and be in a phase
in which $\theta_\rho$ and $\theta_\sigma$ are pinned.

We now identify the time reversal polarization with
\begin{equation}
P_\theta = 2\theta_\sigma/\pi \  {\rm mod} \ 2.
\end{equation}
The apparent dependence of $P_\theta$ on the spin quantization axis
is an artifact of abelian bosonization.  In fact, $P_\theta$
is $SU(2)$ invariant.  This can be seen by noting that global
spin rotations are generated
by $S^z \sim \int dx \partial_x\theta_\sigma$ and $S^\pm \sim \int dx \exp \pm 2i \phi_\sigma$.
The latter obeys $[\theta_\sigma,S^\pm] = \pm\pi S^\pm$, so that
$[P_\theta,S^\pm]=0$.  It can further be seen that even in the
presence of spin nonconserving terms in $V_{so}$ as well as the interaction
terms discussed above, $[P_\theta,H]=0$.
Since
$\Theta P_\theta \Theta^{-1} = -P_\theta \ {\rm mod} \ 2$
there are two distinct possible values for the time reversal polarization:
$\langle P_\theta \rangle = 0$ or $1$.  Thus $P_\theta$ can be used
to classify time reversal invariant insulating states.

Consider a finite system with ends.  We now argue that the value
of $P_\sigma$ determines the presence or absence of Kramers
degenerate states at the ends.  The end of a one dimensional system
at $x=0$
must be characterized by a boundary condition for $\theta_\sigma(x=0)$.
Time reversal symmetry limits the possible values to $\theta_\sigma(x=0) =
n \pi$.  The value of $n$, however, depends on how the lattice is terminated.
First suppose that $n=0$.  Then, when $P_\sigma =1$, the pinning of
$\theta_\sigma$ in the bulk is not consistent with the boundary
condition.  The closest it can be is $\langle
\theta_\sigma\rangle = \pm \pi$.  Thus, near the end
there must be a kink of $\pm \pi$ in $\theta_\sigma$ at the end.  Time
reversal symmetry requires these two possibilities to be degenerate,
so there is a Kramers degeneracy of two at the end. On the other hand,
when $P_\sigma = 0$, the bulk energy
gap is ``consistent" with the boundary condition, allowing for
$\theta_\sigma(x)=0$ everywhere.  The groundstate in this case is
unique.

We thus conclude that bosonization provides an alternative approach
for formulating the time reversal polarization in terms of
$\theta_\sigma$, just as it allows for a formulation of the charge
polarization $P_\rho = \theta_\rho/\pi$.  This suggests
 that the topological distinction of the $Z_2$ spin pump
remains in the presence of electron interactions.

\section{Discussion}

\subsection{Relation to Quantum Spin Hall Effect}

The quantum spin Hall phase introduced in Ref. \onlinecite{km1} is a phase of a two
dimensional electron system.  In a manner analogous to Laughlin's
construction for the quantum Hall effect\cite{laughlin}, this phase, when
compactified onto a cylinder, defines a $Z_2$ pump of the sort
studied in this paper.  In this section we outline the implications
of the present work for the quantum spin Hall effect.  We begin by
relating the $Z_2$ index introduced in section III to the index which
distinguishes the quantum spin Hall phase from a band insulator.  We
then discuss the presence or absence of gapless edge states in the
quantum spin Hall effect.  Finally, we comment on an alternative
topological characterization of the quantum spin Hall effect in terms
of a ``Chern number matrix" that has recently been proposed by Sheng et al.
\cite{sheng}.

\subsubsection{$Z_2$ classification of quantum spin Hall phase}

For non interacting electrons, the electronic phase of a two
dimensional system with a bulk gap is characterized by the
wavefunctions defined on the Brillouin zone torus,
$|u_n(k_x,k_y)\rangle$.  The relationship between the one dimensional
$Z_2$ pump and the two dimensional quantum spin Hall effect can be
established by the identification of $(k,t)$ with $(k_x,k_y)$. Eq.
(\ref{cond2})
then reflects the time reversal invariance of the two dimensional
Hamiltonian.  As we prove in the Appendix, the $Z_2$ topological
index introduced in Ref. \onlinecite{km2} is equivalent to the $Z_2$ index
characterizing the pump. The considerations of this paper provide a
natural physical interpretation of this index in terms of the change
in the time reversal polarization in half of the cycle.

In addition, our observation that the time reversal polarization is
related to the presence or absence of a Kramers degeneracy at the end
suggests that the $Z_2$ classification of time reversal invariant two
dimensional ground states transcends the non interacting model for
which it was derived.  This means that the quantum spin Hall effect
describes a phase that is distinct from a band insulator even in the
presence of electron electron interactions.

\subsubsection{Edge states or not?}

In the regular quantum Hall effect, the topological structure of the
bulk state guarantees the existence of gapless edge excitations. The
non trivial $Z_2$ invariant, however, does {\it not} guarantee
gapless edge states.  As shown by Wu et al.\cite{wu} and Xu et al.\cite{xu},
when the interactions at
the edge are sufficiently strong the edge can undergo a transition
which opens a gap. The considerations of this paper allow us to prove
that {\it either} there are gapless edge excitations {\it or} there
is a ground state degeneracy at the edge associated with the breaking
of time reversal symmetry.

\begin{figure}
 \centerline{ \epsfig{figure=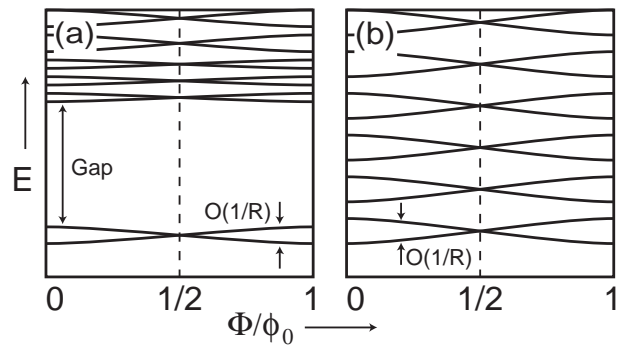,width=3.2in} }
 \caption{Schematic plots of the many particle eigenstates of a cylindrical
 quantum spin Hall system of radius $R$
 as a function of the magnetic flux threading the cylinder.
 In each case there is a Kramers degeneracy when the flux $\Phi$ is
 equal to $h/2e$.  (a)  There exist gapless edge excitations, whose
 energy level spacing goes to zero for $R\rightarrow \infty$.  (b)
 There is an edge excitation gap which remains finite for
 $R\rightarrow\infty$, but the ground state is doubly degenerate.
 }
 \end{figure}

To establish this proposition, consider the many particle eigenstates
of the quantum spin Hall phase on a cylinder as a function of the
magnetic flux through the cylinder.  When the radius $R$ of the
cylinder large, then the $O(1/R)$ change in the energy of the many
particle eigenstates when the one half a flux is inserted will be
much less than any energy gap. But in the quantum spin Hall state,
the non trivial $Z_2$ index requires the ground state to have a
Kramers degeneracy at either $\Phi=0$ or $\Phi = \phi_0/2$, but not
both.  Thus there are two possibilities as schematically illustrated
in Fig. 4. Either there are edge states with energy $O(1/R)$ which
become gapless for $R\rightarrow\infty$ or the ground state is
degenerate for $R\rightarrow\infty$ and split by at most $O(1/R)$ by
the magnetic flux.

This required ground state degeneracy distinguishes the quantum spin
Hall phase from that of a band insulator.  Unlike a band insulator,
the quantum spin Hall state in a system with edges can {\it not} have
a unique ground state with a gap for all excitations.

\subsubsection{Other proposed classifications of the quantum spin
Hall effect}

We now comment on a different topological classification of the
quantum spin Hall effect proposed by Sheng et al. \cite{sheng}
These authors defined a matrix of Chern numbers by
considering a system with a generalized class of periodic boundary
conditions.  Specifically, they considered boundary conditions of the
form $\Phi(...,{\bf r}_{i\alpha}+{\bf L_j},...) = \exp(
i\theta^\alpha_j) \Phi(...,{\bf r}_{i\alpha},...)$,
where ${\bf L}_{j=x,y}$ define the periodicity and
$\theta_j^{\alpha=\uparrow,\downarrow}$ are {\it independent} phase
twists for the up and down spins.
They then characterized the topological classes of the
groundstate wavefunction as a function of these phase twists, and
defined a matrix of Chern numbers, $C^{\alpha,\beta}$.

Sheng et al. argued that this classification contains more
information than the $Z_2$ classification because it distinguishes
quantum spin Hall states which belong to the same $Z_2$ class.  This
can be illustrated by looking at the continuum version of the graphene
model
introduced in Ref. \onlinecite{km1}, described by the Hamiltonian
\begin{equation}
H = \psi^\dagger \left[ -i v_F(\sigma_x\tau_z \partial_x +
\sigma_y\partial_y) + \Delta_{so}\sigma_z\tau_z s_z \right]\psi.
\end{equation}
Here, in the notation of  Ref. \onlinecite{km1}, $\sigma_z$ describes the sublattice of the
honeycomb lattice, $\tau_z$ describes the two inequivalent valleys at
the corners of the Brillouin zone and $s_z$ describes the spin.  When
$\Delta_{so}$ is nonzero the system is in a quantum spin Hall phase
and belongs to the nontrivial $Z_2$ class.  Sheng et al. \cite{sheng} argued that
the {\it sign} of $\Delta_{so}$ defines two distinct phases which are
distinguished by the matrix of Chern numbers.

When $s_z$ is conserved this is certainly correct, and the Chern
number matrix can be viewed as independent Chern numbers for the up
and down spins.  However, when spatial symmetries are relaxed and
spin is not conserved this distinction is no longer meaningful.
The two phases discussed above are in fact the {\it same} phase
because they can be continuously transformed into one another without
closing the gap.  Specifically, consider the more general spin orbit
interaction
which preserves the energy gap:
\begin{equation}
\sigma_z\tau_z s_z \rightarrow \sigma_z \tau_z (\vec s \cdot \hat n).
\label{path}
\end{equation}
When the unit vector $\hat n$ is continuously varied from $+\hat z$
to $-\hat z$ the two ``phases" are connected.  Of course, the process
of connecting these phases requires the breaking of the $C_3$ lattice
symmetry of graphene.  But in general, disorder will break all
spatial symmetries, so one can not rely on a spatial symmetry to
protect a topological property.

This presents a conundrum because the Chern matrix formulation
distinguishes the two states with distinct topological integers - even when
the $C_3$ symmetry is explicitly violated.
What happens to these integers when the continuous path in Eq. (\ref{path}) is
adiabatically followed?   The answer is that somewhere along the
path the energy gap must vanish at the {\it edge} where the twisted
spin boundary condition is imposed\cite{qi}.

The spin phase twist imposed by Sheng et al. can be decomposed into a
$U(1)$ part $\theta_\rho = \theta_\uparrow+\theta_\downarrow$ and
a ``spin" part $\theta_\sigma = \theta_\uparrow-\theta_\downarrow$.
The spin phase twist $\theta_\sigma$ is fundamentally different from $\theta_\rho$
 when the bulk Hamiltonian does not commute with the
$S_z$.  The boundary where the spin phase twist
is imposed is physically different from the rest of the system, and
the spectrum of the Hamiltonian will in general be different for
different values of $\theta_\sigma$.  In contrast, the location of the
charge phase twist $\theta_\rho$ introduced by
Niu and Thouless \cite{niu1}, can be moved around by performing a local gauge
transformation without changing the spectrum.  Since the vector
potential can be spread out over the circumference $2\pi R$ of the torus the
change in the spectrum due changing $\theta_\rho$ can be at most $O(1/R)$.  In
contrast, if the spin phase twist is spread out over the
circumference, changing $\theta_\sigma$ changes the Hamiltonian
 by an amount of order $1$.
The spectrum need not be close to that of the physical Hamiltonian.

We conclude that the additional topological structure implied
by the Chern number matrix is a property of the {\it boundary} where the
twisted phase condition is imposed rather than a property of the bulk
two dimensional phase.  The bulk quantum spin Hall effect is
classified by the $Z_2$ invariant alone.

\subsection{Can the $Z_2$ spin pump pump spin?}

Is the $Z_2$ pump we introduced
a spin pump?   Since an isolated $Z_2$ pump returns to its
original state after two cycles, the simple answer to this question is no.
However, any functioning pump must be connected to reservoirs into which
the pump can pump.   In this section we briefly consider the effect
of connecting the $Z_2$ spin pump to reservoirs.  We conclude that
the $Z_2$ pump {\it does} pump spin, though the spin pumped per cycle is
not quantized.  Moreover, we argue that when the coupling to the
reservoirs is weak the $Z_2$ topological structure of the pump is
essential for a nonzero spin to be pumped.  For stronger coupling,
however, the $Z_2$ structure is not essential.

\begin{figure}
 \centerline{ \epsfig{figure=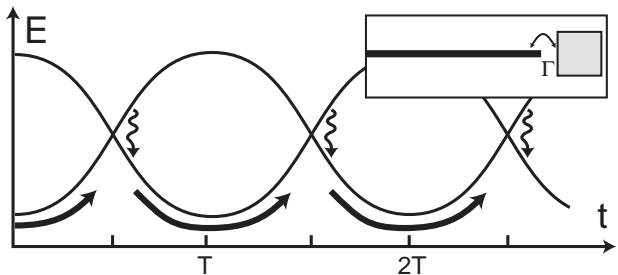,width=3.2in} }
 \caption{Evolution of one end of a
  $Z_2$ pump that is weakly connected to a lead as shown in inset. For $t \gtrsim T/2$
  the excited state of the pump can relax by creating a odd under time
  reversal excitation in the lead, which generically will change the spin
  of the reservoir.}
 \end{figure}

We consider a simple case where the reservoirs can be described by
non interacting electrons with vanishing spin orbit interaction.  We first
suppose the coupling to the reservoir is weak,
so that the level width $\Gamma$ induced in the pump due the coupling is small
compared to the energy gap $\Delta$.  However, we require the coupling $\Gamma$ to
be large compared to the pumping rate, as well as any inelastic
scattering rate for the end states.  In the limit
\begin{equation}
\hbar/T, \hbar/\tau_\varphi \ll \Gamma \ll \Delta,
\end{equation}
the eigenstates of the pump maintain
their integrity, though coupling to the reservoirs allows transitions
between different states.

As illustrated in Fig. 5, there is a point in every cycle $t=(n+1/2)T$,
where the
groundstate of the pump becomes degenerate.  This degeneracy is due to
the end states, which are in proximity to the reservoir.  For $t \gtrsim
(n+1/2)T$ the pump is in an {\it excited} state.  Coupling to the
reservoir, however, allows the pump to relax back to its groundstate.
This relaxation, however, must involve a process in the reservoir which
is {\it odd} under time reversal.  Generically this will
involve changing the spin of the reservoir.  The spin added to the
reservoir need not be quantized.  It could even be equal to zero, but
generically, it will be of order $\hbar$.

In this weak coupling limit it is clear that the $Z_2$ structure of
the pumping cycle is essential because it guarantees the level
crossing in the end states.  If the end states did not cross, then
there would be no transitions, and the spin in the reservoirs would
be unchanged after a complete cycle.  However, finite coupling
between to leads relaxes this requirement.
Suppose that the time reversal symmetry is weakly broken at $t=T/2$,
so that there is a small anticrossing of magnitude $\delta$.
In this case the $Z_2$ character of the cycle is lost.  But if
$\delta \ll \Gamma$, then the states have no way of ``knowing" about
the anticrossing, and the pump proceeds as if $\delta=0$.

This reflects the fact that spin can be introduced into a
reservoir which is connected to an insulating material when the
insulator is deformed through a periodic cycle.  The spin injected
can be expressed in terms of the unitary reflection matrix $\hat r(t)$
for electrons at the Fermi energy in the reservoir\cite{brouwer}, which in general
depends on the Hamiltonian $H(t)$ of the insulator,
\begin{equation}
\Delta \vec S = {1\over{2\pi i}}\oint dt \ {\rm Tr}
[ \vec S \hat r^\dagger {d\hat r\over dt}].
\end{equation}
In general, this quantity is non zero.  The difficulty is coming up
with a cycle in an insulating material for which $\Delta S$ is not
very small.  The $Z_2$ pump accomplishes this by guaranteeing that
there is a resonance in the reflection matrix, which occurs when the
Kramers degenerate end state appears.  Note that this resonance need
not involve charge transfer between the reservoir and the insulator,
for it could be a {\it Kondo resonance} where the Kramers degenerate
end state becomes entangled with the reservoir electrons.

It should be emphasized that the spin added to the reservoir is not a
property of the bulk Hamiltonian of the pump, but rather it depends
on how the pump is connected to the reservoir.  The spin transferred
to the reservoirs at the two ends of the pump need not be related.
Thus, one can not view the spin as being pumped along the
length of the pump.  However, the presence of the end state
resonance, which follows from the change in the time reversal polarization,
{\it is} a property of the bulk insulating state.  In this sense, the
$Z_2$ pump is a pump for spin.


\begin{acknowledgements}

It is a pleasure to thank F.D.M. Haldane, E.J. Mele and T. Pantev for
helpful discussions.  This work was supported by the National Science
Foundation grant DMR-00-79909.

\end{acknowledgements}

\begin{appendix}

\section{Equivalent formulations of the $Z_2$ invariant}

In this appendix we relate different mathematical formulations of the
$Z_2$ invariant $\Delta$.  Our starting point is Eq. (\ref{delta1})
along with Eqs. (\ref{pth1}-\ref{pth4}) which express the invariant in terms of the
change in the time reversal polarization between $t=0$ and $t=T/2$.
We will first show that $\Delta$ can be interpreted as an
obstruction to defining wavefunctions continuously, provided a time
reversal constraint is enforced.  This will lead to a formula for
$\Delta$ in terms of the Berry curvature and Berry connection,
which will be shown to be equivalent to (\ref{delta1}).
We will then show that (\ref{delta1}) is equivalent to the $Z_2$ invariant
proposed for the quantum spin Hall effect in Ref. \onlinecite{km2}.

We will use a notation appropriate
for the $Z_2$ pumping problem and consider Bloch
wavefunctions defined continuously on the torus defined by $-\pi <
k< \pi$ and $0<t<T$.  For the two dimensional quantum spin Hall effect
we should identify $k_x$ with $k$ and $k_y$ with $2\pi t/T$.

\subsection{$Z_2$ invariant as an obstruction}

It is well known that a nonzero value of the Chern invariant is an
obstruction to smoothly defining the wavefunction throughout the
entire torus\cite{kohmoto, nakahara}.  Instead, wavefunctions must be defined on overlapping
``patches" which are related to each other by a gauge transformation called
a ``transition function".  The Chern number is then related to the
winding number of the phase of the transition function around a
non contractable path.    For the problem studied in this paper, the
Chern number is zero, so there is no obstruction to finding a transformation
of the form (\ref{u2n}) which makes the wave
functions smoothly defined on a single patch.  However, we will show in this
section that if we enforce the {\it time reversal constraint}
\begin{eqnarray}
|u^I_\alpha(-k,-t)\rangle = \Theta |u^{II}_\alpha(k,t)\rangle \nonumber\\
|u^{II}_\alpha(-k,-t)\rangle = -\Theta |u^I_\alpha(k,t)\rangle,
\label{tconst}
\end{eqnarray}
then a nonzero value of the $Z_2$ invariant is an
obstruction in a manner precisely analogous to the Chern number.
This constraint means
that the gauges for the wavefunctions at $\pm(k,t)$ are not independent.
At the four time reversal invariant points
$(k,t)=\Gamma_i$, the allowed transformations of the form (\ref{u2n})
are restricted to be {\it symplectic}, $U_{mn}(\Gamma_i) \in Sp(N)$.
That a nonzero value of the $Z_2$ invariant $\Delta$ is inconsistent with this
constraint is easy to see because it implies that ${\rm Det}[w(k,t)]
= 1$ for all $k$ and $t$ and ${\rm Pf}[w(\Gamma_i)] = 1$, so Eq.
(\ref{delta2})
trivially gives $\Delta = 0$.

We will now relate the $Z_2$ invariant to the winding of the
phase of transition functions relating the wavefunctions on different
patches.  In addition to establishing the connection between the
$Z_2$ invariant and the Chern invariant, this approach will derive a
formula for the $Z_2$ invariant which expresses it in terms of
the Berry's connection and curvature.  The similarity between the
$Z_2$ invariant and the Chern invariant has been emphasized by
Haldane\cite{haldaneprivate}.  The formulation of the $Z_2$ invariant as an obstruction
has also been discussed by Roy\cite{roy}, though that work did not establish a
formula for the invariant.

Suppose that we have wavefunctions obeying (\ref{tconst}) defined
smoothly on two patches in the torus labeled A and B in Fig. 6.  In patch A the
wavefunctions $|u^s_\alpha(k,t)\rangle_A$ are smoothly defined everywhere
in the upper left and lower right quadrants of Fig. 6, while for patch B
$|u^s_\alpha(k,t)\rangle_B$ are defined in the upper right and lower
left quadrants.  In the overlapping regions these different
wavefunctions are related by a $U(2N)$ transition matrix
\begin{equation}
|u_m(k,t)\rangle_A = t^{AB}_{mn}|u_n(k,t)\rangle_B.
\end{equation}
where $m$ and $n$ are shorthand of $s$ and $\alpha$.
Consider the change in the $U(1)$ phase of $t^{AB}$ around the
closed loop $\partial\tau_1$ in Fig. 6.
\begin{equation}
D = {1\over{2\pi i}}\oint_{\partial\tau_1}
d\ell \cdot {\rm Tr}[t^{AB\dagger} \nabla t^{AB}].
\label{d1}
\end{equation}
This will clearly be an integer because it is equal to the winding
number of the phase of ${\rm Det}[t^{AB}]$ around the loop
$\partial\tau_1$.
If $D$ is non zero and can not be eliminated by a gauge
transformation, then there is an obstruction to smoothly defining the
wavefunctions on a single patch.  In what follows, we show that
$D\ {\rm mod}\ 2$ is precisely equal
to the $Z_2$ invariant defined in this paper.

\begin{figure}
 \centerline{ \epsfig{figure=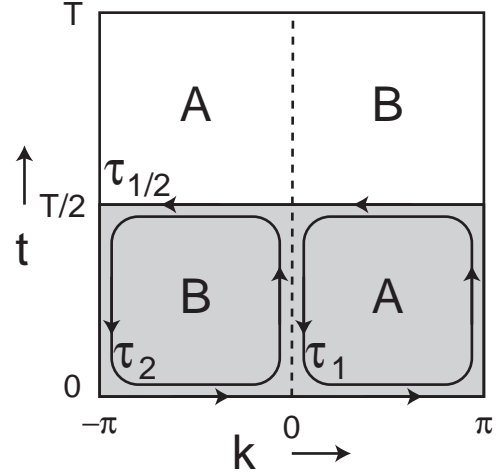,width=2.5in} }
 \caption{The torus defined by $k$ and $t$ divided into two patches A
 and B.  The boundaries of the regions $\tau_1$ and $\tau_2$ are
 shown as well as the boundary of the shaded region, $\tau_{1/2}$.
 }
 \end{figure}

From (\ref{d1}), we may write
\begin{equation}
D = {1\over{2\pi }} \oint_{\partial\tau_1} d\ell \cdot ({\cal A}^B - {\cal
A}^A),
\end{equation}
where  ${\cal A}^A = {\sum_n}  _A\langle
u_n|\nabla|u_n\rangle_A$ and likewise for ${\cal A}^B$.
Since $|u_n\rangle_A$ is smoothly defined in the interior
of $\tau_1$ we may write it in terms of the Berry's flux,
\begin{equation}
\oint_{\partial\tau_1} d\ell \cdot {\cal A}^A = \int_{\tau_1}d\tau {\cal
F}^A.
\end{equation}
This can not be done for $|u_n\rangle_B$ which is not
necessarily defined continuously inside $\tau_1$.  However, it can be
related to the Berry's flux through $\tau_2$,
\begin{eqnarray}
\oint_{\partial\tau_1} d\ell \cdot {\cal A}^B =
- \oint_{\partial\tau_2} d\ell \cdot {\cal A}^B +
\oint_{\partial\tau_{1/2}}d\ell \cdot {\cal A}^B \\
= - \int_{\tau_2} d\tau{\cal F}^B + \oint_{\partial\tau_{1/2}}d\ell
\cdot {\cal A}^B.
\end{eqnarray}
Combining these we thus find the winding number for the transition
function can be expressed as an integral involving the Berry
connection and the Berry curvature.
\begin{equation}
D = {1\over{2\pi}} \left[ \oint_{\partial\tau_{1/2}} d\ell \cdot {\cal
A}- \int_{\tau_{1/2}} d\tau{\cal F}\right] \ {\rm mod} \ 2.
\label{d2}
\end{equation}
The patch labels can be safely removed because ${\cal F}$ is gauge
invariant, and the line integral is gauge invariant modulo 2.
It is essential that the time reversal constraint (\ref{tconst}) be enforced for
this equation to have meaning.  If not, then a gauge transformation
on patch B
can change the line integral by 1, making the formula vacuous.  When
(\ref{tconst}) is enforced, an {\it odd} value of $D$ can not be gauged
away because the phases of $|u^I\rangle$ and $|u^{II}\rangle$ can
not be independently changed.  Thus, $D=1 \ {\rm mod} \ 2$ presents an obstruction to
defining wavefunctions on a single patch.

We now show that this winding number is precisely the same as the
invariant $\Delta$.
To this end, we rewrite $\Delta$ in terms of the {\it partial
polarization} $P^I$ defined in \ref{ps}.  Using $P_\theta = 2 P^I-P_\rho$
we have
\begin{equation}
\Delta = 2 \left(P^I(T/2) - P^I(0)\right) -
\left(P_\rho(T/2)-P_\rho(0)\right) \ {\rm mod} \ 2.
\end{equation}
When (\ref{tconst}) is enforced Eq. (\ref{pi2}) for the partial polarization
implies that
\begin{equation}
2 \left(P^I(T/2) - P^I(0)\right) = {1\over {2\pi}}\oint_{\partial
\tau_{1/2}}
d\ell \cdot {\cal A}.
\end{equation}
Eq. (\ref{prho}) shows that
\begin{equation}
P_\rho(T/2)-P_\rho(0) ={1\over{2\pi}} \int_{\tau_{1/2}}d\tau {\cal F}.
\end{equation}
Combining the two terms thus establishes that $\Delta = D$.  The two
terms in (\ref{d2}) thus acquire physical meaning:  The line integral gives
twice the change in the partial polarization between $t=0$ and $t=T/2$,
while the surface integral gives the change in the total
polarization.

\subsection{Zeros of the Pfaffian}

In Ref. \onlinecite{km2}, the $Z_2$ invariant was introduced by considering the
matrix elements of the time reversal operator,
\begin{equation}
m_{ij}(k,t) = \langle u_i(k,t)|\Theta |u_j(k,t)\rangle.
\end{equation}
This should be contrasted with the matrix $w_{ij}(k,t)$ introduced in section III,
which can be generalized as a function of $t$ to be
\begin{equation}
w_{ij}(k,t) = \langle u_i(-k,-t)|\Theta|u_j(k,t)\rangle.
\end{equation}
At the four time reversal invariant points
$(k,t) =\Gamma_{1,2,3,4} = (0,0),(\pi,0),(0,T/2),(\pi,T/2)$, $w_{ij}$ and
$m_{ij}$ coincide, but in general they are different.
$w_{ij}$ is unitary with $|{\rm Det}[w]|=1$, while  $m_{ij}$ is not
unitary.  Since $\Theta^2=-1$, $m_{ij}$ is antisymmetric.  The Pfaffian of
$m$ is therefore defined for all $k$ and $t$.  In Ref. \onlinecite{km2} we argued
that the $Z_2$ invariant could be determined by counting the number
of zeros of the Pfaffian in half the torus, modulo 2.

To establish the equivalence of this with (\ref{delta1}), we begin by rewriting
the time reversal polarization $P_\theta$ in terms of ${\rm
Pf}[m(k,t)]$.
The key observation to be made is that
\begin{equation}
{\rm Det}[w(k,t)] = {{\rm Pf}[m(k,t)] \over{ {\rm
Pf}[m(-k,-t)]^*}},
\end{equation}
which can be proved by noting that
$m(-k,-t) = w(k,t) m(k,t)^* w(k,t)^T$ and using
the identity ${\rm Pf}[X A X^T] = {\rm Det}[X] {\rm Pf}[A]$.
Introducing $p(k) = {\rm Pf}[m(k,t^*)]$ for $t^*=0,T/2$ it follows that
${\rm log}{\rm Det}[w(k,t^*)] = {\rm log}\ p(k) + {\rm log}\ p(-k)$.
Thus we may rewrite (\ref{pth3}) as
\begin{equation}
P_\theta = {1\over {2\pi i}}\left[ \int_0^\pi dk \nabla_k [{\rm log}\
p(k)+{\rm log}\ p(-k)] - 2 {\rm log} \left ({p(\pi) \over p(0)} \right)
\right],
\label{pth5}
\end{equation}
where we have used the coincidence of $w$ and $m$ at $k=0$ and $\pi$.
This may be simplified further by changing variables $k \rightarrow -k$ in
the middle term and writing
the last term as an integral from $0$ to $\pi$.  This gives
\begin{equation}
P_\theta = {1\over {2\pi i}} \int_{-\pi}^\pi dk \nabla_k {\rm log} {\rm Pf}[m(k,0)] \ \ {\rm mod}\
2,
\end{equation}
where the integral is now over the closed loop $t=0$, $-\pi<k<\pi$.
This expression is only defined modulo 2 because of the ambiguity of
the imaginary part of the log in (\ref{pth5}).

Thus, we have established that
$P_\theta$ is given by the phase winding of the
Pfaffian, $p(k)$ around the 1D Brillouin zone modulo 2.  While this quantity
is not gauge invariant, the change in it due to continuous evolution
between $t^*=0$ and $t^*=T/2$ is gauge invariant.   This defines the
$Z_2$ topological invariant, which, as in Ref. \onlinecite{km2}, may be written
\begin{equation}
\Delta = {1\over {2\pi i}} \oint_{\partial \tau_{1/2}} d\ell\cdot\nabla {\rm log} {\rm Pf}[m(k,t)]
\ \ {\rm mod}\ 2,
\end{equation}
where $\partial\tau_{1/2}$ is the boundary of half the torus defined
by $-\pi<k<\pi$ and $0<t<T/2$ (see Fig. 6).
If ${\rm Pf}[m(k,t)]$ has point zeros, then this quantity counts the
number of zeros in $\tau_{1/2}$ modulo 2.

\end{appendix}

\end{document}